\documentstyle[amstex,amssymb,aps,preprint,12pt]{revtex}
\begin{document}
\title{Excitation of Triple Giant Resonances in Heavy-Ion Reactions$^{\ast }$}
\author{M.S.~Hussein$^{1}$, B.V.~Carlson$^{2}$, L.F. Canto$^{3}$, and A.F.R. de
Toledo Piza$^{1}$}
\address{$^{1}$Instituto de F\'{i}sica, Universidade de S\~{a}o Paulo,\\
CP 66318, 05315-970 S\~{a}o Paulo, Brazil\\
$^{2}$Departamento de F\'{i}sica, Instituto Tecnol\'{o}gico de
Aeron\'{a}utica - CTA\\
12228-900 S\~{a}o Jos\'{e} dos Campos, SP, Brazil\\
$^{3}$Instituto de F\'{i}sica, Universidade do Rio de Janeiro,\\
CP 68528, 21945-970 Rio de Janeiro RJ, Brazil}
\date{\today}
\maketitle

\begin{abstract}
We calculate the cross-section for the excitation and subsequent decay of
triple giant resonances (TGDR) in several nuclei excited with heavy ions. The
recently developed coherent plus incoherent theory for the excitation in
conjunction with the hybrid decay model of Dias-Hussein-Adhikari are used
for the purpose. It is emphasized that the direct decay of the TGR is
expected to deviate appreciably from the harmonic limit especially at low
bombarding energies, owing to the incoherent contribution.
\end{abstract}
PACS Numbers: 25.70.De, 24.30.Cz, 21.10.Re 

\vspace{1cm}

$\ast $Supported in part by CNPq and FAPESP.\newpage

The study of the double giant dipole resonance in nuclei has received a
considerable amount of attention over the last 15 years [1]. Both the pion
double charge exchange and relativistic heavy ion Coulomb excitation
reactions have been used to probe this large amplitude collective motion in
may fermion systems. The quest for the similar double plasmon resonance in
metallic clusters is underway [2]. Plans are also in progress to search
for the triple giant dipole resonance (TGDR) in nuclei [3]. It is clearly of
importance to supply theoretical estimates of the cross-section as well as
the different decay branching ratios of these exotic collective modes. This
is the purpose of the present paper. We use the recently developed coherent
plus incoherent excitation theory of Ref. [4] in conjunction with the hybrid
decay model of Dias-Hussein-Adhikari (DHA) of Ref. [5].

The existing models for the calculation of the excitation cross-section of
DGDR can be grouped into three categories: a microscopic structure model in
conjunction with second order Coulomb excitation perturbation theory [6], a
macroscopic, oscillator (harmonic or anharmonic) model in conjunction with
coupled channels Coulomb excitation theory [7] and finally the recently
developed average plus fluctuation model [4,8]. In this latter model the
average cross-section is calculated according to the theory developed in
[9], where the simple, double, etc. giant resonances are considered as
doorway states belonging to the spectrum of a damped harmonic oscillator.

A fully microscopic structure calculation of the excitation cross-section of
the TGDR resonance is prohibitively difficult. A huge number of three
particle-three hole configurations have to be dealt with in a coupled
channel context. A detailed account of the spreading of the TGDR would
require the inclusion in the calculation of at least the four particle -
four hole subspace. Not having available such a detailed description we opt
for using our coherent + fluctuation model [8]. The TGDR excitation
cross-section is found to have the form

\begin{equation}
\sigma ^{(3)}=\sigma _{c}^{(3)}+\sigma_{fl}^{(3)}(2) +\sigma _{fl}^{(3)}(1)\,
\end{equation}

\noindent where $\sigma _{c}^{(3)}$ is the average
cross-section for the coherent excitation of the three phonons which
proceeds through the one and two phonon states in a typical three-step
description. This cross section can be guaranteed for the coupled channel
Coulomb excitation model which contains explicit reference to flux loss from
there states owing to their spreading into more complex configurations.
The cross-section $\sigma _{fl}^{(3)}(2) $
corresponds to fluctuation contribution arising from the decay of an
intermediate collective GR phonon into the complicated background followed
by the excitation of another collective phonon on the background states
(the Brink-Axel phonon). The final states involved in this cross-section
contains \underline{two} collective phonons in contrast to $\sigma_{c}^{(3)}$.
Finally $\sigma_{fl}^{(3)}(1) $ contains contributions that lead
to only one phonon in the final state. In terms of the time sequence of
events, $\sigma _{c}^{(3)}$, being a three-step process, is the
fastest, followed by the four-step process accounted for by
$\sigma_{fl}^{(3)}(2)$ (this is a four-step process since besides
the three excitation steps one has one internal mixing step) and finally the
five step process contained in $\sigma _{fl}^{(3)}(1) $. The DGR cross section
may be similarly decomposed as
$\sigma^{(2)}=\sigma_c^{(2)}+\sigma_{fl}^{(2)}(1)$.

In Ref. [8] we obtained the following estimates for $\sigma _{fl}^{(3)}( i) $:

\begin{equation}
\sigma _{fl}^{(3)}(2) =\left( \frac{2}{3}\right) 
\frac{\Gamma _{1}^{\downarrow }\text{ }\tau _{c}}{\hslash }\,\sigma
_{c}^{(3)}\,,
\end{equation}

\begin{equation}
\sigma _{fl}^{(3)}(1) =\left( \frac{1}{3}\right)
\left( \frac{\Gamma _{1}^{\downarrow }\,\tau _{c}}{\hslash }\right)
^{2}\text{ }\sigma _{c}^{(3)}\,,
\end{equation}

\noindent where $\Gamma _{1}^{\downarrow }$ is the spreading width of the
single phonon $GR$ and $\tau _{c}$ is the average collision time given by 
$\frac{b_{o}}{\gamma v}$, with $b_{o}$ being the grazing impact parameter, $v$
the asymptotic relative velocity and $\gamma $ is the Lorentz factor,
$\gamma =\left( 1-\left( \frac{v}{c}\right) ^{2}\right) ^{-1/2}$.

The above fluctuation contribution become insignificant at very
high energies where the systems proceeds very quickly though the sequence
ground state $\rightarrow GDR\rightarrow DGDR\rightarrow TGDR$. When the
bombarding energy is lowered, the fluctuation effects may become appreciable
[8]. The estimates above were found to be quite reasonable when compared to
the more elaborate model of Ref. [4] at relatively high bombarding energies. 
At lower energies, clearly, for the evaluation of the different contributions
to the total excitation cross-section, Eq. (1), one should rely on the latter,
more precise model. In order to discuss the decay of the final states into the
open channels, we need to know the values of the different contributions to 
$\sigma^{(2)}$ and $\sigma^{(3)}$.

We have calculated the excitation cross-sections, $\sigma^{(1)}$,
$\sigma^{(2)}$ and $\sigma^{(3)}$, for various nuclei incident on
$^{208}$Pb at several bombarding energies, using a
three-di\-men\-si\-o\-nal (3D) generalization of the model of Ref. [4]. The
3D time evolution equation used to describe the excitation and decay of the
GDR phonons possesses the same form as the one-dimensional equation of
Ref. [4]. However, the collective and statistical excited states of the 3D
model take into account all possible combinations of the (two) transverse
and (one) longitudinal degrees of freedom, which yield 3 coherent one-phonon
states, 6 coherent two-phonon states and 10 coherent three-phonon states as
well as a multitude of states containing a mixture of coherent and
statistical excitations. Decays of the three types of phonons to the
statistical background are assumed to occur independently but to each obey
Bose-Einstein statistics.

The Coulomb interaction matrix elements used to describe the transverse
modes of the GDR excitation in the 3D model are the physically appropriate
ones, as given in Ref. [9]. The longitudinal Coulomb interaction matrix
element, however, is modified from the form given there. It is reduced to a
term proportional to the longitudinal component of the eletric field, in
analogy to the transerse terms, but which differs from the expression given
in Ref. [9] by a total time derivative. The latter term can be extracted
from the equations and discarded when only the coherent excitation is
included. This is no longer the case when decay to the statistical states is
taken into account. Nevertheless, we have neglected its contribution here.

As in Ref. [4], the coupled equations of motion are solved as a function of
impact parameter to yield asymptotic occupation probabilities. Effective
asymptotic occupation probabilities are defined, for states that decay, as
the sum over the probability that decays out of each state during the time
evolution. Cross sections are obtained by integrating each probability x
differential area over impact parameter and summing over polarizations.

The various contributions to the cross sections are easily extracted from the
theoretical calculations. In Table I, we present the coherent and fluctuation
contributions to the DGDR cross section, $\sigma _{c}^{(2)}$ and
$\sigma_{fl}^{(2)}(1)$ for various nuclei incident on $^{208}$Pb at several 
energies. In Table II, we present the contributions to the TGDR cross section,
$\sigma _{c}^{(3)}$, $\sigma _{fl}^{(3)}(2)$, and $\sigma _{fl}^{(3)}(1)$.
We observe that the cross sections increase
dramatically with the charge of the projectile. As is well known, the
coherent two-phonon cross sections scales approximately as the charge Z
squared, while the three-phonon one scales as $Z^{3}$. We also observe that
the coherent contribution to the cross sections only dominates at relatively
high incident energies. At the lowest energy shown, 10 MeV/nucleon, the
cross sections involving statistical phonons are substantially larger than
the coherent ones. At $E/A=100$ MeV, it is clear from the tables that the
fluctuation contribution to the DGDR cross section is about as large as the
coherent one, while the fluctuation contribution to the TGDR is about three
times larger than the coherent contribution. 

We turn now to the decay of the DGR and TGR. We first remind the reader of
the hybrid direct+fluctuation decay model of DHA [5]. According to this model,
which has been extensively used in the analysis of the decay data [10,11],
the GR decays to a find channel $f$ in the following manner:

\begin{eqnarray}
\sigma^{(1)}_{f}&=&\sigma^{(1)}\left[ \left( 1-\mu _{1}\right) 
\frac{\tau _{f}^{(GR)}}{\sum_j\,\tau _{j}^{(GR)}}+ 
\mu _{1}\frac{\tau _{f}^{(CN)}+\mu _{1}\tau _{f}^{(GR)}}{\sum_j\,
\left( \tau _{j}^{(CN)}+\mu _{1}\tau _{j}^{(GR)}\right) }\right] \nonumber \\
&\equiv& \sigma^{(1)}\,(P_f^{\uparrow}+P_f^{\downarrow})\,,\label{grdec}
\end{eqnarray}
where $\sigma^{(1)}$ is the one phonon excitation cross section 
discussed before, while 
$\mu _{1}=\frac{\Gamma _{1}^{\downarrow }}{\Gamma _{1}}$ and the 
$\tau$'$s$ are the appropriate transmission coefficients.
We have written the probability of populating the final channel $f$ through
direct decay of the GR as
\begin{equation}
P_f^{\uparrow}=\left( 1-\mu _{1}\right) 
\frac{\tau _{f}^{(GR)}}{\sum_j\,\tau _{j}^{(GR)}}\,,
\end{equation}
and the probability of of populating the channel $f$ through the statistical
states as
\begin{equation}
P_f^{\downarrow}=\mu _{1}\frac{\tau _{f}^{(CN)}+\mu _{1}\tau _{f}^{(GR)}}
{\sum_j\,
\left( \tau _{j}^{(CN)}+\mu _{1}\tau _{j}^{(GR)}\right) }
\end{equation}
Note that the statistical decay component contains explicit reference to
the GR direct transmission, $\left( \mu _{1}\tau _{f}^{(GR)}\right)$. 

Before entering into the details of the decay of the multiple giant
resonances, let us
first analyze the decomposition into direct decay and decay into the
statistical states. For this purpose, we use the branching ratios
\begin{equation}
P^{\uparrow}=\sum_f\,P^{\uparrow}_f = 1-\mu_1\qquad\qquad\mbox{and}
\qquad\qquad P^{\downarrow}=\sum_f\,P^{\downarrow}_f=\mu_1\,.
\end{equation}
The decomposition of the single GR decay is a direct result of 
Eq.~(\ref{grdec}),
\begin{equation}
\sigma^{(1)}\rightarrow (P^{\uparrow}+P^{\downarrow})\,\sigma^{(1)}\,.
\end{equation}

To decompose the decay of the multiple giant resonances into direct and
statistical parts, we assume that each of the collective phonons decays
independently. The decay of the coherent contributions to the DGR and TGR
can then be decomposed as 
\begin{eqnarray}
\sigma_c^{(2)}&\rightarrow& ( P^{\uparrow}+P^{\downarrow})^2\, \sigma_c^{(2)} 
\label{coh2} \\
&=&( P^{\uparrow\,2} +2 P^{\uparrow}P^{\downarrow} +P^{\downarrow\,2})\,
\sigma_c^{(2)}\,,\nonumber
\end{eqnarray}
and
\begin{eqnarray}
\sigma_c^{(3)}&\rightarrow& ( P^{\uparrow}+P^{\downarrow})^3\, \sigma_c^{(3)} 
\label{coh3} \\
&=&( P^{\uparrow\,3} +3 P^{\uparrow\,2} P^{\downarrow}
+3 P^{\uparrow}P^{\downarrow\,2} +P^{\downarrow\,3})\,
\sigma_c^{(3)}\,.\nonumber
\end{eqnarray}
That is, the coherent contribution to the DGR can decay through direct 
decay of each of the collective phonons, through a direct decay of one of the
colective phonons and decay into the statistical states of the other, or
through decay into the statistical states of both of the phonons. Decay of the
coherent contribution to the TGR takes into account the different possible
direct or statistical decays of the three initial phonons. 

We can analyze the decomposition of the decay of the fluctuating contributions
to the DGR and TGR cross sections in a similar manner. We need only take into
account the number of collective phonons in each of the contributions. Thus,
the decay of the fluctuating component of the DGR can be decomposed as
\begin{equation}
\sigma_{fl}^{(2)}(1)\rightarrow (P^{\uparrow}+P^{\downarrow})\,
\sigma_{fl}^{(2)}(1)\,,
\end{equation}
while the decay of the fluctuating components of the TGR can be decomposed as
\begin{equation}
\sigma_{fl}^{(3)}(2)\rightarrow ( P^{\uparrow\,2} 
+2 P^{\uparrow}P^{\downarrow} +P^{\downarrow\,2})\,\sigma_{fl}^{(3)}(2)\,,
\end{equation}
and
\begin{equation}
\sigma_{fl}^{(3)}(1)\rightarrow (P^{\uparrow}+P^{\downarrow})\,
\sigma_{fl}^{(3)}(1)\,.
\end{equation}

We can now combine the various terms into decompositions of the decay of the
total DGR and TGR cross sections, 
\begin{eqnarray}
\sigma^{(2)}&\rightarrow& P^{\uparrow\,2}\,\sigma_c^{(2)} \label{all2} \\
 &+&
P^{\uparrow}\,(2 P^{\downarrow}\,\sigma_c^{(2)}+\sigma_{fl}^{(2)}(1))
\nonumber \\
 &+& P^{\downarrow}\,( P^{\downarrow}\,\sigma_c^{(2)}+\sigma_{fl}^{(2)}(1))\,,
\nonumber
\end{eqnarray}
and
\begin{eqnarray}
\sigma^{(3)}&\rightarrow&  P^{\uparrow\,3}\, \sigma_c^{(3)} \label{all3} \\
 &+& P^{\uparrow\,2}\,(3 P^{\downarrow}\,\sigma_c^{(3)}+\sigma_{fl}^{(3)}(2)) 
\nonumber \\
 &+& P^{\uparrow} \,(3 P^{\downarrow\,2}\,\sigma_c^{(3)}
+2 P^{\downarrow}\,\sigma_{fl}^{(3)}(2)+\sigma_{fl}^{(3)}(1)) \nonumber \\
 &+& P^{\downarrow} \,(P^{\downarrow\,2\,}\sigma_c^{(3)}
+ P^{\downarrow}\,\sigma_{fl}^{(3)}(2)+\sigma_{fl}^{(3)}(1))\,,\nonumber
\end{eqnarray}
where we have collected terms according to the number of direct decays
involved.

In medium to heavy nuclei, one expects the spreading to dominate over the
escape from the GR, which implies $P^{\downarrow}=\mu _{1}\simeq 1$.
In this case, we see that the completely statistical DGR and TGR decay cross
sections, given by the last term in the preceding two equations, will be
approximately proportional to their total excitation cross sections. They will
not distinguish between the coherent and fluctuating components of the cross
sections. Further analysis of the decay of the statistical component to
equilibrium can be quite complicated. Usually, however, particle emission from
the statistical component can be well described using the equilibrium 
Hauser-Feshbach formalism.

To best view the distinction between the coherent and fluctuation
contributions to the GDR cross sections, we look for effects in the direct
decay from the giant resonance. We have already analyzed the effects of the
coherent and fluctuating components on exclusive decays of the
DGDR.\cite{ref12}. Here, we wish to analyze their effects on inclusive
decay cross sections of the DGR and TGR. 

To separate direct contributions to an inclusive cross section from the
statistical contributions, it is necessary to concentrate one's attention
on the high-energy end of the emission spectrum. There, the statistical
weight of the high energy residuals in the statistical cross section
strongly suppresses emission, leaving the direct emission to dominate the
spectrum. To obtain the inclusive emission cross section from our
decomposition of the decay of the excitation cross sections through direct and
statistical modes, Eqs.~\ref{all2} and \ref{all3}, we take into account that
each of the direct decay factors $P^{\uparrow}$ can contribute independently
to direct emission in channel $f$ with a relative probability
$P_f^{\uparrow}/P^{\uparrow}$. Thus, for example, each of the factors of 
$P^{\uparrow}$ in the $P^{\uparrow\,2}$ (first) term of Eq.~\ref{all2} 
contributes to the direct emission in channel $f$ with probability  
$P_f^{\uparrow}/P^{\uparrow}$, resulting in a total contribution of 
$2P_f^{\uparrow}/P^{\uparrow}$ of the term to the emission cross section.
In general, we include a factor of
$P_f^{\uparrow}/P^{\uparrow}$ for each of the factors of $P^{\uparrow}$
appearing in the equations. We neglect the terms containing only statistical
decays, since we have $P_f^{\downarrow}/P^{\downarrow}\approx 0$ at the
high end of the emission spectrum. 

We point out that the relative decay probabilities 
$P_f^{\uparrow}/P^{\uparrow}$ are not quit the same for each of the decays, 
since conservation laws constrain the energies and angular momenta of
the emitted particles to values consistent with the residual values in the
nucleus. The residual nuclear values will not be identical for a cold
nucleus (first emission) and a hot one (second and later emissions).
Nevertheless, the general structure of the phonon, which we assume to be
essentially a one particle-one hole state, assure that the value of
$P_f^{\uparrow}/P^{\uparrow}$ will be about the same for each of the emissions.

Using the rules above for estimating the contribution of the $P^{\uparrow}$
factors to emission, we find the direct inclusive emission spectra to be
given by
\begin{eqnarray}
\sigma_f^{(2)}&=& 2 \frac{P_f^{\uparrow}}{P^{\uparrow}}\,
P^{\uparrow\,2}\,\sigma_c^{(2)}  \\
 &+&\frac{P_f^{\uparrow}}{P^{\uparrow}}\,
P^{\uparrow}\,(2 P^{\downarrow}\,\sigma_c^{(2)}+\sigma_{fl}^{(2)}(1))\,,
\nonumber 
\end{eqnarray}
and
\begin{eqnarray}
\sigma_f^{(3)}&=& 3\frac{P_f^{\uparrow}}{P^{\uparrow}}\,
 P^{\uparrow\,3} \,\sigma_c^{(3)} \\
 &+& 2\frac{P_f^{\uparrow}}{P^{\uparrow}}\,P^{\uparrow\,2} 
(3 P^{\downarrow}\,\sigma_c^{(3)}+\sigma_{fl}^{(3)}(2)) \nonumber \\
 &+& \frac{P_f^{\uparrow}}{P^{\uparrow}}\,P^{\uparrow} 
(3 P^{\downarrow\,2}\,\sigma_c^{(3)}
+2 P^{\downarrow}\,\sigma_{fl}^{(3)}(2)+\sigma_{fl}^{(3)}(1))\,, \nonumber 
\end{eqnarray}
which reduce to
\begin{equation}
\sigma_f^{(2)} = P_f^{\uparrow}\,(2 \sigma_c^{(2)} + \sigma_{fl}^{(2)}(1))\,,
\label{inc2}
\end{equation}
and
\begin{equation}
\sigma_f^{(3)} = P_f^{\uparrow}\,(3 \sigma_c^{(3)} + 2 \sigma_{fl}^{(3)}(2)
+\sigma_{fl}^{(3)}(1))\,.\label{inc3}
\end{equation}
Here we see clearly the importance of the (more) coherent contributions to the 
direct emission spectra. Each of the excitation cross sections contributes
according to the number of collective phonons it possesses.

We can similarly analyze the contribution to the inclusive cross sections of
each of the components of the of the excitation cross sections. In particular,
we determine the contribution of the coherent DGR and TGR to the emission
spectra to be
\begin{equation}
\sigma_{c,f}^{(2)} =2 P_f^{\uparrow}\, \sigma_c^{(2)}\qquad\qquad\mbox{and}
\qquad\qquad\sigma_{c,f}^{(3)} =3 P_f^{\uparrow}\, \sigma_c^{(3)}\,.
\end{equation}
We could call these values the harmonic limit of the cross section. Comparing
these to the values for the total direct emission, we find
\begin{equation}
\frac{\sigma_f^{(2)}}{\sigma_{c,f}^{(2)}} = 1  +\frac{1}{2}
\frac{\sigma_{fl}^{(2)}(1)}{\sigma_{c}^{(2)}}\,,
\end{equation}
and
\begin{equation}
\frac{\sigma_f^{(3)}}{\sigma_{c,f}^{(3)}} = 1  + \frac{2}{3} 
\frac{\sigma_{fl}^{(3)}(2)}{\sigma_{c}^{(3)}}
+\frac{1}{3}\frac{\sigma_{fl}^{(3)}(1)}{\sigma_{c}^{(3)}}\,.
\end{equation}
Thus, a considerably larger direct decay may occur if the fluctuation
contributions are important, which may occur at lower bombarding energies.
Of course, one could obtain deviation of the direct decay from the harmonic
limit (two or three independently decaying phonons), if anharmonic effects were
allowed, This, however, will imply deviation of the spectrum of the
oscillator from the harmonic sequence, which seems to be borne out
neither by experiment [1] nor by calculation [7].

In conclusion, we have, in this paper, calculated the excitation
cross-section and studied the the decay properties of the double and
triple giant dipole resonances of various nuclei as excited in Coulomb
collisions with $^{208}$Pb. It was found that the
degree of deviation of the direct decay from the limit of two or three
independently decaying collective phonons depends significantly on the
bombarding energy, $E$, and can be appreciable at low values of $E$.

\newpage

\begin{table}
% {\bf Two Phonon Cross Sections}
\caption{
Contributions of the coherent and fluctuation components to the DGDR
excitation cross section (in mb) of various projectiles incident on a
lead target at several different values of the incident energy.}
\begin{tabular}{|c|rr|rr|rr|}
\hline
& \multicolumn{2}{c|}{10 MeV} & \multicolumn{2}{c|}{100 MeV} & 
\multicolumn{2}{c|}{1 GeV} \\ 
Projectile & $\sigma_{c}^{(2)}$ & $\sigma_{fl}^{(2)}(1)$ & $\sigma_{c}^{(2)}$
& $\sigma_{fl}^{(2)}(1)$ & $\sigma_{c}^{(2)}$ & $\sigma_{fl}^{(2)}(1)$ \\ 
\hline
$^{40}$Ca & 1.01 & 4.52 & 2.17 & 2.19 & 7.20 & 0.72 \\ 
$^{120}$Sn & 12.37 & 49.54 & 26.48 & 22.94 & 72.61 & 6.65 \\ 
$^{132}$Xe & 15.03 & 59.62 & 32.19 & 27.57 & 88.50 & 8.00 \\ 
$^{165}$Ho & 23.32 & 90.61 & 51.13 & 42.60 & 138.59 & 12.34 \\ 
$^{208}$Pb & 40.45 & 148.66 & 96.95 & 72.87 & 234.84 & 19.83 \\ 
$^{238}$U & 47.05 & 174.78 & 109.15 & 84.86 & 276.53 & 24.04 \\ \hline
\end{tabular}
\end{table}

\vspace{2.0cm}

\begin{table}
%  {\bf Three Phonon Cross Sections}
\caption{
Contributions of the coherent and fluctuation components to the TGDR
excitation cross section (in mb) of various projectiles incident on a
lead target at several different values of the incident energy.}
\begin{tabular}{|c|rrr|rrr|rrr|}
\hline
& \multicolumn{3}{c|}{10 MeV} & \multicolumn{3}{c|}{100 MeV} & 
\multicolumn{3}{c|}{1 GeV} \\ 
Projectile & $\sigma_{c}^{(3)}$ & $\sigma_{fl}^{(3)}(2)$ & $%
\sigma_{fl}^{(3)}(1)$ & $\sigma_{c}^{(3)}$ & $\sigma_{fl}^{(3)}(2)$ & $%
\sigma_{fl}^{(3)}(1)$ & $\sigma_{c}^{(3)}$ & $\sigma_{fl}^{(3)}(2)$ & $%
\sigma_{fl}^{(3)}(1)$ \\ \hline
$^{40}$Ca & 0.01 & 0.08 & 0.11 & 0.02 & 0.06 & 0.02 & 0.11 & 0.02 & 0.00 \\ 
$^{120}$Sn & 0.17 & 2.46 & 3.17 & 0.84 & 1.92 & 0.64 & 3.03 & 0.47 & 0.04 \\ 
$^{132}$Xe & 0.23 & 3.20 & 4.10 & 1.10 & 2.50 & 0.83 & 4.07 & 0.62 & 0.05 \\ 
$^{165}$Ho & 0.42 & 5.81 & 7.37 & 2.08 & 4.70 & 1.54 & 7.76 & 1.17 & 0.09 \\ 
$^{208}$Pb & 0.95 & 12.24 & 14.83 & 5.28 & 10.78 & 3.36 & 16.68 & 2.40 & 0.18
\\ 
$^{238}$U & 1.18 & 15.09 & 18.60 & 6.13 & 13.14 & 4.22 & 21.01 & 3.14 & 0.24
\\ \hline
\end{tabular}
\end{table}

\end{document}